\documentstyle[12pt,a4,amssymb,epsfig]{article}

\def\l{\left(}
\def\r{\right)}

\newcommand{\be}{\begin{equation}}
\newcommand{\ee}{\end{equation}}

\renewcommand{\ln}{\mathop{\rm ln}\nolimits}

\newcommand{\bg}{\begin{gather}}
\newcommand{\eg}{\end{gather}}

\begin{document}
\begin{center}
{\Large\bf Small Second Acoustic Peak from Interacting Cold Dark Matter?}\\
\vspace{0.3cm}
 S.~L.~Dubovsky\footnote{{\bf e-mail}: sergd@ms2.inr.ac.ru}, 
D.~S.~Gorbunov\footnote{{\bf e-mail}: gorby@ms2.inr.ac.ru} \\
{\small{\em 
Institute for Nuclear Research of the Russian Academy of Sciences, }}\\
{\small{\em
60th October Anniversary prospect 7a, Moscow 117312, Russia
}}
\end{center}
\begin{abstract} 
   We consider a possibility to explain the observed suppression of
the second acoustic peak in the anisotropy spectrum of the Cosmic
Microwave Background (CMB) by interaction between a fraction of
non-baryonic Cold Dark Matter (CDM) and normal baryonic matter.  This
scenario does not require any modifications in the standard Big Bang
Nucleosynthesis (BBN). We estimate the required values
of the cross-section-to-mass ratio for elastic
scattering of CDM particles off baryons. 
In case of velocity-independent elastic
scattering (in the velocity interval $\upsilon\sim10^{-5}\div10^{-3}$) 
we find that such particles do
not contradict observational limits if they are heavier than $\sim
10^5$~GeV or lighter than $\sim 0.5$~GeV. 
Another candidate, which may appear
in the models with infinite extra dimensions, is a quasistable charged
particle decaying through tunneling into extra dimensions.
Finally a millicharged particle with the
electric charge ranging from $\sim 10^{-4}$ to $\sim 10^{-1}$ and with
mass $M\sim 0.1\mbox{~GeV}\div 1\mbox{~TeV}$ also may be responsible 
for the suppression of the second acoustic peak. As a byproduct we point
out that CMB measurements set new limits on the allowed parameter
space for the millicharged particles.
\end{abstract} 

Recently the anisotropy spectrum of the CMB has been measured with great
precision resulting in the accurate determination of the shape and
position of the first acoustic peak~\cite{BOOMERanG&MAXIMA}. The
result, being in a good agreement with the standard inflationary
predictions such as adiabatic spectrum of
primordial fluctuations and flat Universe, 
provides a  strong observational support to the inflationary
picture of the Early Universe. 

Current data include the values of angular harmonics
up to $l\sim 800$ covering the region where the second acoustic
peak has been expected. In the standard inflationary
$\Lambda$CDM Universe, the relative height of the first and the second
acoustic peaks is governed by the fractional baryon mass density
$\Omega_Bh^2$ (see Ref. \cite{Silk} for introduction to 
physics of the CMB anisotropy and further references). The latter
parameter is independently determined from the primordial element 
abundances, $\Omega_Bh^2=0.019$, with an accuracy of about
5\%~\cite{BBN}. The CMB anisotropy measurements suggest somewhat higher
baryon density~\cite{jaffe}, 
\begin{equation}
\Omega_Bh^2=0.032^{+0.005}_{-0.004}
\label{omega-b}
\end{equation} 
(the upper limit here is valid assuming the prediction of
simplest inflationary models that the
tensor to scalar ratio is small, $r\approx0$ \cite{deviation})
which deviates from the standard BBN value at about 2$\sigma$
level. New CMB measurements should reduce the
uncertainty in the result~(\ref{omega-b}) 
in the near future and thus allow to
determine whether this discrepancy is real or just a
statistical fluctuation. 

Consequently, it is desirable to have a 
list of physical effects capable to modifying the CMB
spectrum in the region of first acoustic peaks. 
When experimental uncertainties are reduced, these
effects either will help to explain the potential discrepancy or will
be ruled out by the CMB measurements.

There are two broad classes of existing proposals aimed
to resolve the above discrepancy. The first approach
is to take the high value of $\Omega_B$ as granted by the CMB
anisotropy measurements and try to make the predictions of BBN compatible
with it. This can be
achieved either by relaxing some of the assumptions of the standard
BBN scenario (see, e.g., \cite{BBNs}) or assuming reduction of photon 
entropy between nucleosynthesis and recombination epoch
\cite{turner}. Another approach is to find a mechanism of suppression
of the second acoustic peak. Such mechanism 
may invoke (see,
e.g. \cite{tilt}), for
instance, a tilted  spectrum of primordial fluctuations 
and large reionization optical depth, leading to
damping of the features in the CMB spectrum at small angles.

It is important to note that the approach based on the high baryon
density (``baryon drag'') predicts 
the suppression of all even peaks relative to the odd
ones \cite{Silk}. 
Future precise measurements of the third acoustic peak are likely to 
discriminate between the two approaches. 

In this note we discuss a mechanism of the third type which also leads
to the suppression of all even acoustic peaks but does not require high
baryon density and, as a result, is compatible with the standard BBN
scenario.  In order to describe our mechanism, let us, following
Ref.~\cite{Silk},
briefly recall
the physics of acoustic oscillations and how the baryon drag works. 

When the size of the primordial adiabatic density fluctuation becomes 
smaller than the horizon scale, its amplitude starts to grow due to
the gravitational instability. At the epoch of interest (roughly,
between radiation--matter equality and recombination) the
primordial matter has two components ---
photon-baryon-electron plasma and CDM. 
The pressureless CDM component experiences the
gravitational infall providing seeds for the formation of the structure in
the Universe, while the density perturbations in the plasma cease to
grow and turn into the acoustic oscillations, as the pressure due to 
the photon component compensates for the gravitational
attraction. After recombination, when photons and baryons become
non-interacting, baryons fall in the gravitational wells formed by the
CDM density fluctuations giving rise to galaxies, while photons
propagate freely. Inflation
provides equal initial amplitudes for the oscillations at 
different scales, so the density contrast in the acoustic wave (and,
as a result, photon temperature) at recombination is determined by
its phase only, which in turn is determined by the ratio of the wavelength
of the oscillation to the sound horizon at recombination. This gives
rise to the oscillatory structure in the angular spectrum of CMB. 

Higher baryon density leads to reduction of pressure in the plasma, 
that changes the balance between pressure and gravity and shifts the
zero point of the oscillations. This results in enhancement of the
amplitudes of all odd peaks relative to the even ones. Then it is
clear that the same effect could come from any other massive particles
($X$-particles) constituting a fraction of CDM and interacting with
the components of the plasma strongly enough to be involved in the
acoustic oscillations. Certainly, this condition implies that
$X$-particles cannot be responsible for the seeds for the galaxy
formation, so they cannot be the only component of the CDM. On the
other hand, the difference between values of $\Omega_Bh^2$ evaluated 
from BBN and the CMB measurements is about $\sim (1/2)\Omega_B$. 
Consequently, the density $\Omega_X$ of the
$X$-particles as small as a few percent of the total density is 
sufficient to make these two predictions compatible with each other, so
there is still enough room for the conventional CDM, weakly
interacting with the photon-baryon plasma.
  
In this note we estimate the suitable range of parameters (masses
and interactions) of the $X$-particles and discuss existing limits on 
various candidates. 

Let us consider $X$-particles with mass 
$M_X$, which 
can scatter elastically off protons with the 
cross section $\sigma_{Xp}$. The rate of the 
energy transfer from the photon-baryon plasma to these particles
is
\be
\label{*}
{dE_X\over dt}=\Delta E_X\tau^{-1} 
\ee
where $\Delta E_X$ is the energy transfer per one collision and $\tau$
is the characteristic time between the collisions. 
Assuming isotropic scattering one has
\be
\Delta E_X=2{m_p M_X\over (m_p+M_X)^2}(E_p-E_X)\;,
\ee
where $E_p$ and $E_X$ are kinetic energies of the proton and
$X$-particle, respectively. 
The timescale $\tau$ between two subsequent collisions is
\be
\tau=(n_B \sigma_{Xp}\upsilon)^{-1}\;,
\label{tau}
\ee
where $n_B$ is the baryon number density and $\upsilon$ is relative 
velocity. Protons heat $X$-particles in the collisions. Due to this process
$X$-particles may come into kinetic equilibrium with protons with
corresponding time scale $t_{eq}$ given by 
\be
\label{4*}
t^{-1}_{eq}={2\over 3T}{dE_x\over dt}\;,
\ee
where $T$ is the temperature of the plasma. $X$-particles will be
involved in the acoustic oscillations at recombination and will suppress
the amplitude of the second acoustic peak if the equilibration time
$t_{eq}$ does not exceed the inverse expansion rate of the Universe
$H_r^{-1}$ at that moment. Combining Eqs.~(\ref{*})--(\ref{4*}) 
and expressing the Hubble
constant $H_r$ and baryon density $m_pn_B$ through their present 
values $H_0$ and $\Omega_B\rho_c$ 
one evaluates the following bound for the cross-section-to-mass ratio
of the $X$-particle 
\[
{\sigma_{Xp}M_X^{1/2}\over (M_X+m_p)^{3/2}}\gtrsim
{H_0\over 2\Omega_B}\sqrt{m_p\over 3T_r}{1\over\rho_{c}}
\l {T_0\over T_r}\r^{3/2}\;,
\]
where $T_0=2.7$~K and $T_r=0.25$~eV are the present CMB temperature
and the temperature at recombination.  Taking the standard BBN value
$\Omega_Bh^2=0.02$ one obtains
\begin{equation}
\sigma_{Xp}
\gtrsim2.7h\cdot 10^{-22}~{(M_X+m_p)^{3/2}\over M_X^{1/2}}{{\rm cm}^2
\over {\rm GeV}}\;.
\label{sigma-over-mx}
\end{equation}
Here $\sigma_{Xp}$ is the cross section corresponding to 
the relative velocity $\sim10^{-5}$. 
Certainly, this lower bound implies extremely large cross sections. 
In concrete models one can evaluate $\sigma_{Xp}(\upsilon)$ and find
the interval of $M_X$ allowed by present experimental data. 

As the first example, let us consider the case in which 
the cross section $\sigma_{Xp}$ is velocity-independent for
$\upsilon\sim10^{-5}\div10^{-3}$. Then the 
cross section of $X$ scattering off baryons in halo is also larger
than the right hand side of Eq.~(\ref{sigma-over-mx}). 
Surprisingly, as it was emphasized
recently~\cite{spikes-2}, even if all CDM species interact equally
strongly with ordinary
baryonic matter, as large cross section as
\begin{equation}
\sigma_{Xp}=8\cdot 10^{-25}\div
1\cdot10^{-23}\left({M_X\over {\rm GeV}}\right)~{\rm cm}^2\;.
\label{spikes-sigma-over-mx}
\end{equation} 
at the relative velocity $\sim 10^{-3}$ 
cannot be excluded at present if CDM particles are heavier than
$\sim 10^5$~GeV or lighter than $\sim 0.5$~GeV. 
The bound~(\ref{sigma-over-mx}) requires an 
order of magnitude higher cross section for the same mass. However the
upper bound in Eq.~(\ref{spikes-sigma-over-mx}) is not
applicable in our case. Indeed this bound comes from the consideration of
the heating rate $\gamma$ of non-ionized interstellar clouds by elastic
collisions of $X$-particles with Hydrogen~\cite{cooling}. Namely, this
limit was
obtained by requiring that the heating rate $\gamma$ is smaller than
the observed cooling rate $\lambda=(8.1\pm 4.8)\times
10^{-14}$~eV/s~\cite{coolobs}.  Upper limit in
Eq.~(\ref{spikes-sigma-over-mx}) assumes that $X$-particles constitute
all Dark Matter in the halo. However, as we noted above, as low
density of $X$-particles as $\Omega_Xh^2=0.01$, is sufficient to suppress the
amplitude of the second acoustic peak to the observed
value. Correspondingly, the heating rate $\gamma$, which is
proportional to the density of $X$-particles, is suppressed at the
same values of $M_X$ and $\sigma_{Xp}$ by a factor of 
\[
\Omega_X/\Omega_M\sim 1/15\div 1/25
\]
for $\Omega_M=0.3\div 0.5$~. 
 Consequently, the upper bound is weaker, by this factor, as compared
 to Eq.~(\ref{spikes-sigma-over-mx}) and
does not contradict Eq.~(\ref{sigma-over-mx}). 

Another upper limit
on $\sigma_{Xp}$ close to that in
Eq.~(\ref{spikes-sigma-over-mx}) is related to the halo stability
 \cite{spikes-2} and 
does not apply to $X$-particles for the same reason --- we do not
assume that $X$-particles constitute the halos of galaxies, so the
fact that they are stopped in the disc does not 
lead to the instability of the halos.

Finally, it is worth noting that the interval of parameters given by
Eq. (\ref{spikes-sigma-over-mx}) is interesting also because
self-interaction of the CDM particles in this range may resolve the
problem of the weakly interacting CDM model, predicting overly dense
cores in the centers of galaxies and clusters and an overly large
number of galaxies within the Local Group in contradiction with
observations~\cite{spikes-1}.  It is intriguing that this interval is
not only very close to the parameters of hadronic interactions as was
stressed in Ref.~\cite{spikes-2} but to the bound
(\ref{sigma-over-mx}) as well.

To summarize, a relatively small fraction of CDM, consisting of 
any stable particles with mass $M_X\lesssim 0.5$~GeV (a lower bound 
depends on the concrete model of the interaction with baryons) or 
$M_X>10^5$~GeV which elastically 
scatter off baryons
with the velocity-independent (at $\upsilon\sim10^{-5}\div10^{-3}$) 
cross section $\sigma_{Xp}$, obeying Eq.~(\ref{sigma-over-mx}),
guarantees the suppression of the second acoustic peak in the anisotropy
spectrum of CMB and does not contradict to the present observational limits. 

Another natural candidate to consider is the
 electrically charged massive particles
(champs)~\cite{glashow,champ2}. 
First, let us consider particles with unit charge $X^+$ and $X^-$. 
As was explained in Ref.~\cite{glashow} the fates of these particles in
the Early Universe are very different. Positive champs survive
unscathed till the epoch of recombination when they capture electrons
and form superheavy Hydrogen. Consequently, before recombination they
can be involved in the acoustic oscillations. This happens if their
equilibration time in plasma is smaller than the inverse expansion
rate $H_r^{-1}$ at recombination. The equilibration time for
$X^+$ in plasma is given by \cite{glashow}
\begin{equation}
\label{equilibration}
t_{eq}={3M_Xm_p\over 8\sqrt{2\pi}q^2\alpha^2 n_B \ln\left(3T/q\alpha
k_D\right)}\left({T \over m_p}+{T\over M_X}\right)^{3/2}\;,
\end{equation}
where $q=1$ is the electric charge of $X^+$ and $k_D=(4\pi
n_e\alpha/T)^{1/2}$ is the Debye momentum. This time is smaller than
the expansion rate of the Universe at recombination provided
$M_X<10^{11}$~GeV. Consequently, these particles could lead to the
suppression of the second acoustic peak.

Negative champs can form electromagnetic
bound states with proton and nuclei. In fact, as it was argued in
Ref.~\cite{glashow}, the dominant part of $X^-$
form neutral bound states with proton (``neutrachamps''). 
The cross section of the elastic scattering of neutrachamps off protons at
small relative velocities $v_p\lesssim 10^{-3}$
was estimated in Ref.~\cite{glashow} by making use of the results for
the scattering of slow ions off neutral atoms, 
\begin{equation}
\label{neutrachamp}
\sigma_{pX}=0.36 {\pi^2\over m_p^2\upsilon_p^2}\;.
\label{champ-proton}
\end{equation}
The interval of $M_X$, where
inequality~(\ref{sigma-over-mx}) is valid, is forbidden by searches
for massive particles in isotopes (for recent review and corresponding
references see Ref.~\cite{review-ijmp}). So neutrachamps 
cannot be responsible for the suppression of the
second acoustic peak but could form seeds for the galaxies.

Unfortunately, as pointed out in Ref.~\cite{graveyard}, this
scenario is plagued with observational difficulties. Namely, if champs
constituted a significant part of the matter in halo, some of the
superheavy hydrogen formed by the positive champs would be trapped by
the protostellar clouds during the star formation.  Neutrachamps are
not captured by these clouds, so $X^+$ have no chance to annihilate
in the stars and will live there indefinitely. If the evolution of the
host star leads to the creation of a neutron star, $X^{+}$ will form
a black hole in the center of the neutron star and will destroy it on 
timescales much shorter than the life-time of the Universe.  This
argument rules out champs with masses up to $10^{16}$~GeV constituting
a significant fraction  of the halo.

One can try to get around this argument in two different ways. First, one
can imagine that there exists a pair of nearly degenerate particle
species in which the
heavier one is charged and the lighter one is neutral. If the lifetime
of the charged particle is somewhat larger than the age of the
Universe at recombination and the mass difference between charged and
neutral particles is small enough so that the late decay of the former
does not lead to the strong distortion of the photon background, then
such particles can be responsible for the suppression of the second
acoustic peak. Certainly any model of this type requires strong 
fine-tuning. 

Unusual modification of this scenario may be realized
in models with infinite extra dimensions where our world is localized on a 
3-brane in a non-compact multidimensional space~\cite{RS}-\cite{LRS}. 
Namely, as shown 
in Ref.~\cite{DRT1}, it is possible that
massive (even charged!) particles have a finite probability to 
escape into extra 
dimensions. From the point of view of the 4-dimensional observer such
"decay" would mean a literal dissapearance of the particle. 
In Refs.~\cite{GRS,DRT2}
it was shown that this proccess is compatible with the Gauss
laws of the general relativity and electrodynamics and 
leads to the dissapearence of the gravitational (electric) field of the 
particle in the causal way. As the dissappeance of the particles into extra 
dimensions is a tunneling process, the corresponding 
life-time is naturally very large.
Its value depends on the particular mechanism of localization,
fundamental parameters of the underlying multidimensional theory and mass of 
the particle. The simplest case is the scalar field localized 
on the brane by the gravitational field.  In this case the decay rate is 
given by \cite{DRT1}
\be
\label{decay}
\Gamma={\pi M\over \Gamma{\l 2+n/2\r}\Gamma{\l1+n/2\r}}\l{M\over 2k}\r^{2+n}~,
\ee
where $M$ is a mass of the particle, $(n+1)$ is the number of extra dimensions
(one of them is infinite and $n$ are compact and warped, 
see Ref.~\cite{DRT2}), $k$ is the inverse AdS radius in the bulk. 
Assuming that this scale is of the order of 4-dimensional Planck mass,
$k\sim 10^{19}$~GeV, one has the following value for the ratio
of the life-time of the scalar
particle $t_s$ to the life-time of the Universe at the recombination 
$t_r\sim 1.8\cdot 10^{13}$~sec,
\be
\label{lifetime}
t_s/t_r={5\cdot 10^{-58}\over \Gamma{\l 2+n/2\r}\Gamma{\l1+n/2\r}} \l 
{2\cdot 10^{19}~{\rm GeV}\over M} \r^{3+n}\;.
\ee
It follows from Eq.~(\ref{lifetime}) that for $n=0$, 
in order to survive untill the recombination, 
the scalar particle should have mass smaller than a few GeV which is
not allowed by collider searches if the particle is
charged\footnote{It is worth noting also that in the simple model
which we consider gauge fields are gravitationaly localized on the
brane only at $n>0$ \cite{DRT2}.}. 
However, for larger values of $n$, a life-time of the particle can be in 
the interesting range (larger than the age of the Universe 
at recombination and
smaller than about 1 Gyr when first stars were formed, 
in order to avoid the neutron
star argument) without strong fine-tuning of the parameters. 
For instance, for $n=1$ the mass of the particle should be in the range
\[
20~\mbox{TeV}\lesssim M\lesssim 160~\mbox{TeV}\;.
\]
 
Another way to get around the neutron star argument is to consider
particles of smaller charge. Fluxes of the particles of 
charges $q\gtrsim 0.1$ are strongly limited 
by direct searches \cite{Review-ijmp}, so
one has to consider particles with smaller charges, $q\lesssim
0.1$. Let us check that particles with so small electric charges can
give rise to the suppression of the second acoustic peak. Comparing
equilibration time (\ref{equilibration}) 
for $X$-particle of charge $q$ with the inverse expansion rate
of the Universe at recombination one obtains that the $X$-particle 
mass should be less than
\begin{equation}
\label{millimass}
M_X<3.4\cdot10^9q^2(18.6-\ln q)\mbox{~GeV}
\end{equation}
for $M_X>m_p$, and larger than
\begin{equation}
\label{millimass1}
M_X>7.3\cdot10^{-20}q^{-4}(18.6-\ln q)^{-2}\mbox{~GeV}
\end{equation}
for $M_X<m_p$, 
in order that $X$-particles have been involved in the acoustic oscillations. 
\begin{figure}[htb]
\begin{center}
\epsfig{file=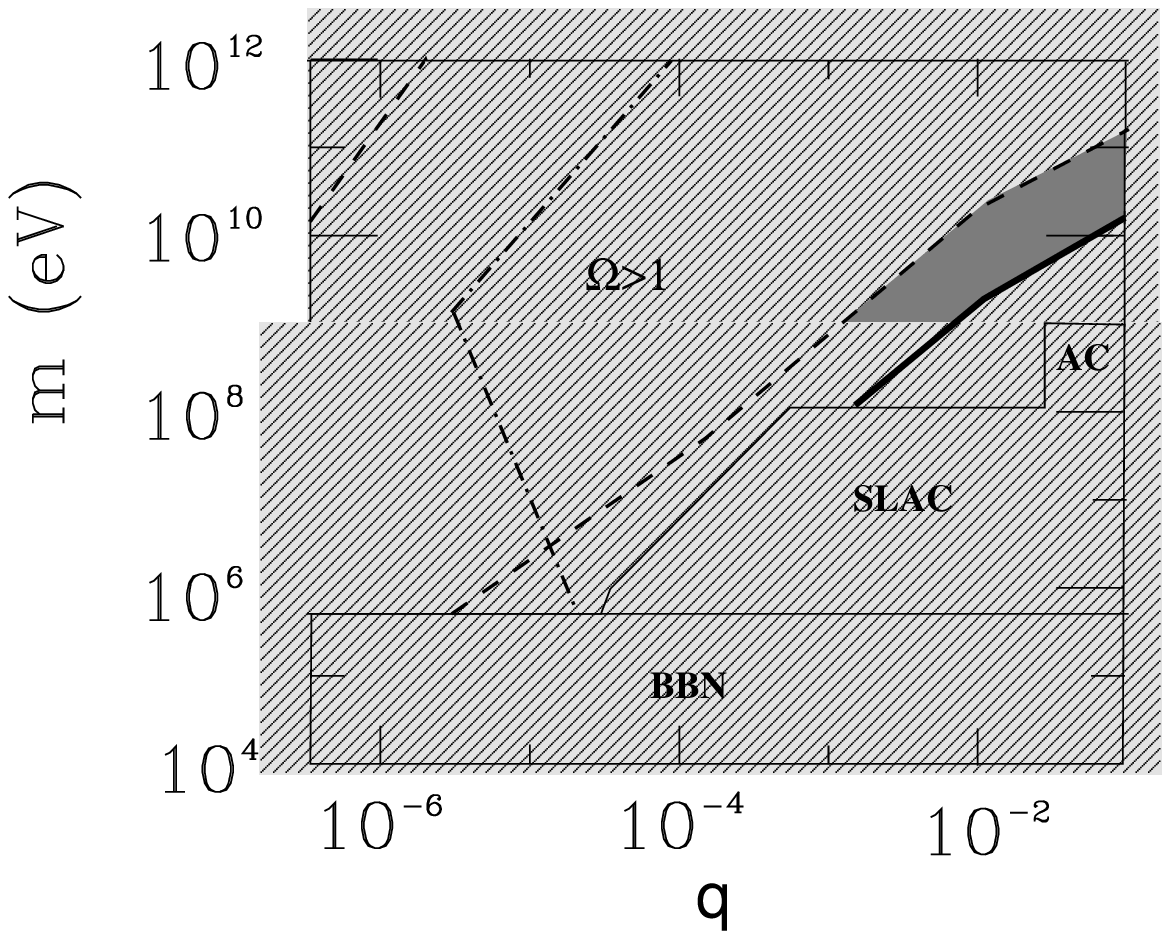}
\end{center}
\caption{An exclusion plot in the mass-charge space for millicharged
particles in models without paraphotons. 
Dashed lines correspond to the relic density of
millicharged particles equal to $\Omega_X h^2=1$.  To the right of 
dash-dotted
line millicharged particles are involved in the acoustic
oscillations. Light grey shaded area is experimentally
excluded. Models with fractional charge $q\gtrsim0.1$ are ruled out by
strong limits on fluxes of the millicharged particles. 
Dark grey area is excluded, if millicharged particles
do not contribute to the oscillatory structure of angular CMB
spectrum. Thick solid line corresponds to models explaining the
absence of second acoustic peak.}
\label{plot}
\end{figure}
In Fig. \ref{plot} we present this bound (dash-dotted line) in the
exclusion plot for the parameters of models with millicharged
particles and without paraphotons (see Ref. \cite{millicharge}). 
There exists allowed region of parameters where
millicharged particles are involved in the acosutic oscillations. 
Dark grey area corresponds to the region of allowed parameter space
where masses of millicharged particles satisfy Eq.~(\ref{millimass})
and their relic density is higher than $\Omega_{mc}h^2=0.01$. The CMB 
result~(\ref{omega-b}) implies that this
region of parameter space is excluded (assuming inflationary spectrum 
with $r=0$ and baryon density favored by the standard BBN). In
particular we practically close the window for 
light millicharged particles with
charges $10^{-5}\div 10^{-3}$. Millicharged particles of masses and
charges belonging to the thick
solid line in Fig.~\ref{plot} may be responsible for the observed
suppression of the second acoustic peak.

In models with paraphotons the CMB 
result~(\ref{omega-b}) implies that region with
$q\sim10^{-4}\div10^{-1}$ and $M_X\sim~a~few$~TeV is excluded 
(assuming inflationary spectrum with 
$r=0$ and baryon density as in the standard BBN scenario). 
Millicharged particles with $M_X\approx1$~TeV and
$q\approx10^{-4}\div10^{-1}$ may be responsible for the observed
suppression of the second acoustic peak.

To conclude, the observed suppression of the second
acoustic peak in the CMB anisotropy may be due to the interaction
of the primordial photon-baryon plasma with $X$ particles constituting
a fraction of the CDM of order a few percent. In the case of elastic
velocity-independent scattering (in the velocity interval
$\upsilon\sim10^{-5}\div10^{-3}$) the required cross sections
(\ref{sigma-over-mx}) are extremely large, but intriguingly similar to
the recently predicted range (\ref{spikes-sigma-over-mx}) for the
self-interaction of the CDM.  In this case the direct searches for Dark
Matter restrict $X$-particles to be within mass 
interval $\sim 0.1\div 0.5$~GeV or
to be heavier than $10^5$~GeV. Another 
candidate which may be responsible for the 
suppression of the second acoustic peak emerges 
in multidimensional scenarios
with non-compact extra dimensions. It is a charged 
particle, stable in the multidimensional theory, which may 
escape from our brane through 
tunneling into extra dimensions. Yet another possibility is the
existance of millicharged particles with the electric charge ranging
from $\sim 10^{-4}$ to $\sim 10^{-1}$ and with mass $M_X\sim
0.1\mbox{~GeV}\div 1\mbox{~TeV}$. 

The authors are indebted to V.A.~Kuzmin, V.A.~Rubakov, D.V.~Semikoz,
M.E.~Shaposhnikov,  P.G.~Tinyakov and I.I.~Tkachev 
for helpful discussions. SD thanks
Institute of Theoretical Physics, University of Lausanne, where this
work was completed, for hospitality. This work was supported in part
by RFBR grant 99-01-18410, by the Council for Presidential Grants and
State Support of Leading Scientific Schools, grant 00-15-96626, by
CRDF grant (award RP1-2103) and by Swiss Science Foundation grant
7SUPJ062239.

\def\ijmp#1#2#3{{\it Int. Jour. Mod. Phys. }{\bf #1~} (19#2) #3}
\def\pl#1#2#3{{\it Phys. Lett. }{\bf B#1~} (19#2) #3}
\def\zp#1#2#3{{\it Z. Phys. }{\bf C#1~} (19#2) #3}
\def\prl#1#2#3{{\it Phys. Rev. Lett. }{\bf #1~} (19#2) #3}
\def\rmp#1#2#3{{\it Rev. Mod. Phys. }{\bf #1~} (19#2) #3}
\def\prep#1#2#3{{\it Phys. Rep. }{\bf #1~} (19#2) #3}
\def\pr#1#2#3{{\it Phys. Rev. }{\bf D#1~} (19#2) #3}
\def\np#1#2#3{{\it Nucl. Phys. }{\bf B#1~} (19#2) #3}
\def\mpl#1#2#3{{\it Mod. Phys. Lett. }{\bf A#1~} (19#2) #3}
\def\arnps#1#2#3{{\it Annu. Rev. Nucl. Part. Sci. }{\bf #1~} (19#2) #3}
\def\sjnp#1#2#3{{\it Sov. J. Nucl. Phys. }{\bf #1~} (19#2) #3}
\def\jetp#1#2#3{{\it JETP Lett. }{\bf #1~} (19#2) #3}
\def\app#1#2#3{{\it Acta Phys. Polon. }{\bf #1~} (19#2) #3}
\def\rnc#1#2#3{{\it Riv. Nuovo Cim. }{\bf #1~} (19#2) #3}
\def\ap#1#2#3{{\it Ann. Phys. }{\bf #1~} (19#2) #3}
\def\ptp#1#2#3{{\it Prog. Theor. Phys. }{\bf #1~} (19#2) #3}
\def\spu#1#2#3{{\it Sov. Phys. Usp.}{\bf #1~} (19#2) #3}
\def\apj#1#2#3{{\it Ap. J.}{\bf #1~} (19#2) #3}
\def\epj#1#2#3{{\it Eur.\ Phys.\ J. }{\bf C#1~} (19#2) #3}
\def\pu#1#2#3{{\it Phys.-Usp. }{\bf #1~} (19#2) #3}
\def\nc#1#2#3{{\it Nuovo Cim. }{\bf A#1~} (19#2) #3}
\end{document}